\documentclass[aps,preprint]{revtex4}

\usepackage{amsfonts}
\usepackage{amsmath}
\usepackage{amssymb}
\usepackage{graphicx}

\begin{document}

\title{Active-sterile neutrino oscillations and pulsar kicks }
\author{M. Barkovich, J. C. D'Olivo, and R. Montemayor}
\thanks{Permanent address: Instituto Balseiro and CAB, Universidad Nacional
de Cuyo and CNEA, 8400 Bariloche, Argentina }
\affiliation{Departamento de F\'{\i}sica de Altas Energ\'{\i}as, Instituto de Ciencias
Nucleares, Universidad Nacional Aut\'{o}noma de M\'{e}xico, Apartado Postal
70-543, 04510 M\'{e}xico, Distrito Federal, M\'{e}xico.}

\begin{abstract}
We develop a thorough description of neutrino oscillations in a
magnetized protoneutron star, based on a resonance layer for
neutrinos with different momentum directions. We apply our
approach to the calculation of the asymmetry in the neutrino
emission during the birth of a neutron star and the pulsar
acceleration in the case of an active-sterile neutrino resonant
conversion. The observed velocities can be obtained with the
magnetic fields expected in the interior of a protoneutron star,
for sterile neutrino masses of the order of $KeV$ and small
mixing angles.
\end{abstract}

\maketitle

\section{Introduction}

The peculiar motion of pulsars is one of the challenging problems in modern
astrophysics. Observations show that young pulsars move with velocities of $%
200-500$ $km\ s^{-1}$ on average, much larger than the mean velocity of
ordinary stars in our Galaxy ($\sim 30$ $km\ s^{-1}$)\cite{pulsar}. This is
strong evidence that pulsars receive a kick during their birth in supernovae
explosions, but the physical origin of such an impulse remains unclear. A
natal kick could be a result of global hydrodynamical perturbations in the
pre-collapse core of the progenitor star or an anisotropic neutrino emission
induced by the strong magnetic fields in the protoneutron star\cite{LCC}. A
significant fraction of the pulsar population has velocities over $1000$ $%
km\ s^{-1}$ which are difficult to account for assuming hydrodynamical kick
mechanisms, as numerical simulations of supernova explosions indicate.

More than $99\%$ of the gravitational energy released in the collapse is
emitted in the form of neutrinos and antineutrinos. A small asymmetry ($\sim
1\%$ ) in the momentum taken away by neutrinos could produce an acceleration
consistent with the measured pulsar velocities. A particle physics
explanation of the origin of this asymmetry has been proposed by Kusenko and
Segr\`{e} (KS)\cite{KS}\cite{otros}, based on adiabatic flavor ($\nu
_{e}\longleftrightarrow \nu _{\mu ,\tau })$ oscillations in matter, in the
presence of a magnetic field. In a magnetized stellar plasma there are
anisotropic contributions to the neutrino refraction index\cite{DNEC}-\cite%
{DN}, and thus the resonant transformation of neutrinos moving in different
directions respect to the magnetic field occurs at different depths within
the protostar, producing a momentum flux asymmetry. With the standard
(active) neutrinos, the mechanism works when the resonance region lies
between the surface of the two neutrinospheres, but it requires an
exceedingly high square mass difference ($\Delta m^{2}\gtrsim $ $10^{4}\
eV^{2}$), in conflict with the existing limits for standard neutrinos.

A variant of the idea of KS can be implemented in terms of matter neutrino
oscillations between active and non interacting (sterile) states \cite{KSS},
in which case the above mentioned limitation can be avoided. More recently
it has been shown that an asymmetric off-resonant emission of sterile
neutrinos $\nu _{s}$ is also a feasible mechanism\cite{KS1}. In both cases,
the range of required oscillation parameters (mass and mixing angle) overlap
with the allowed region for sterile neutrinos considered as the cosmological
dark matter. Alternative neutrino oscillation mechanisms for the pulsar kick
have been proposed in relation to the possible existence of transition
magnetic moments\cite{momen}, non orthonormality of the flavor neutrinos\cite%
{valle}, or violations of the equivalence principle\cite{vep}. The
asymmetric neutrino emission could also produce a short gamma-ray burst\cite%
{kusemi} and gravity waves, whose detection can help to understand pulsar
kicks\cite{grav}.

In a previous article\cite{BDMZ} we reexamined the viability of the KS
mechanism for resonant active-active neutrino oscillations, and emphasized
the relevance of the geometrical ($1/r^{2}$) dependence of the energy flux
to produce an anisotropic neutrino emission. Following previous works on the
subject, we use the concept of an induced deformation by the magnetic field
of an averaged resonance surface, which acts as an effective $\nu _{\mu
,\tau }$ neutrinosphere. In this work we analyze the active-sterile
transformations incorporating a more realistic description in terms of a
spherical layer where the resonant conversion of neutrinos with momentum $k$
moving in all directions takes place. A geometrical expression for the
neutrino flux anisotropy is derived in terms of the ratio of the thickness
of the resonance shell, which depends on the intensity of the magnetic field
and its average position. To quantify the effect we apply our results to two
simple models for the protostar atmosphere.

The paper is organized as follows. In Section II we analyze the resonant
conversion of active neutrinos into sterile neutrinos within the protostar
atmosphere in the presence of a magnetic field, and we derive an expression
for the thickness of the resonance region as a function of the magnetic
field and the medium properties. In Section III, on the basis of the
reasonable assumption that the neutrino flux is in the diffusive regime, we
evaluate the asymmetry of the momentum flux taken out by neutrinos. In
Section IV concrete results are obtained in the context of two specific
models. Section V contains some comments and concluding remarks.

\section{Resonant active-sterile transformation in a magnetized medium}

Within the neutrinosphere neutrinos are trapped due to the opacity of the
medium. This region can be considered as limited by a spherical surface with
a radius $R_{\nu _{l}}$, defined by the condition that above the surface
neutrinos suffer less than one collision\cite{wilson}%
\begin{equation}
\int_{R_{\nu l}}^{\infty }\frac{dr}{\lambda _{\nu _{l}}}\simeq \frac{2}{3},
\label{nc}
\end{equation}%
where $\lambda _{\nu _{l}}$ is the mean free path of the neutrino of flavor $%
l=e,\mu ,\tau $. We consider the protostar atmosphere as constituted by a
degenerate gas of relativistic electrons and a classical nonrelativistic gas
of nucleons. The main contributions to the neutrino opacity in such a medium
come from the neutral-current scattering reactions\cite{B}%
\begin{eqnarray}
\nu _{l}+n &\rightleftarrows &\nu _{l}+n,  \label{r1} \\
\nu _{l}+p &\rightleftarrows &\nu _{l}+p,  \label{r2}
\end{eqnarray}%
and the charged-current absorption reaction%
\begin{equation}
\nu _{e}+n\rightleftarrows e^{-}+p.  \label{r4}
\end{equation}%
The cross sections for the reactions (\ref{r1}) and (\ref{r2}) are%
\begin{eqnarray}
\sigma _{n} &=&\frac{G_{F}^{2}}{4\pi }\left( 1+3g_{A}^{2}\right) k^{2}, \\
\sigma _{p} &=&\sigma _{n}\left[ 1-\frac{8\sin ^{2}\theta _{W}}{1+3g_{A}^{2}}%
\left( 1-2\sin ^{2}\theta _{W}\right) \right] ,
\end{eqnarray}%
where $k$ is the neutrino momentum, $\sin ^{2}\theta _{W}\simeq 0.23$, and $%
g_{A}\simeq 1.26$ is the renormalization of the axial-vector current of the
nucleons. For the absorption reaction we have
\begin{equation}
\sigma _{abs}=4\sigma _{n}\left( 1+\frac{Q}{k}\right) \left[ \left( 1+\frac{Q%
}{k}\right) ^{2}-\left( \frac{m_{e}}{k}\right) ^{2}\right] ^{1/2},
\end{equation}%
with $Q=m_{n}-m_{p}$ denoting the difference between the neutron and the
proton masses.

In terms of thermally averaged cross sections, the mean free paths for the
electron and muon (tau) neutrino are given by
\begin{eqnarray}
\lambda _{\nu _{e}}^{-1} &=&N_{n}\left( \langle \sigma _{abs}\rangle
+\langle \sigma _{n}\rangle \right) +N_{p}\langle \sigma _{p}\rangle \equiv
\kappa _{\nu _{e}}k^{2}\rho ,  \label{l1} \\
\lambda _{\nu _{\mu ,\tau }}^{-1} &=&N_{n}\langle \sigma _{n}\rangle
+N_{p}\langle \sigma _{p}\rangle \equiv \kappa _{\nu _{\mu ,\tau }}k^{2}\rho
,  \label{l2}
\end{eqnarray}%
where $N_{n,p}$ are the neutron and proton number densities, $\rho $ is the
baryonic density, and $\kappa _{\nu _{l}}$ are the neutrino opacities. In a
protoneutron star $N_{n}\simeq 0.9\ \left( N_{n}+N_{p}\right) $ and $k\gg Q$%
, which yields
\begin{equation}
\lambda _{\nu _{e}}^{-1}\simeq 5\lambda _{\nu _{\mu ,\tau }}^{-1}\simeq
5N_{n}\langle \sigma _{n}\rangle ,  \label{ln}
\end{equation}%
and
\begin{eqnarray}
\kappa _{\nu _{e}} &=&3\times 10^{-25}\ MeV^{-5}, \\
\kappa _{\nu _{\mu ,\tau }} &=&0.7\times 10^{-25}\ MeV^{-5}.
\end{eqnarray}%
Eq. (\ref{ln}) implies that $R_{\nu _{e}}>R_{\nu _{\mu ,\tau }}$, as can
immediately be seen from Eq. (\ref{nc}).

In the presence of a magnetic field, the effective potential for an active
neutrino propagating in a plasma of electrons, nucleons, and neutrinos can
be written as\cite{DNEC}\cite{sem}:
\begin{equation}
V_{\nu _{l}}=b_{\nu _{l}}-ec_{\nu _{_{l}}}\ \mathbf{\hat{k}}\cdot \mathbf{B\
,}  \label{v2}
\end{equation}%
with the coefficients $b_{\nu _{l}}$ and $c_{\nu _{_{l}}}$ given by\cite{DN}%
\begin{align}
b_{\nu _{e}}& =\sqrt{2}G_{F}\left[ N_{e}+\left( \frac{1}{2}-2\sin ^{2}\theta
_{W}\right) (N_{p}-N_{e})-\frac{1}{2}N_{n}\right] \text{ }+\tilde{b}_{\nu
_{e}}, \\
b_{\nu _{_{\mu ,\tau }}}& =\sqrt{2}G_{F}\left[ \left( \frac{1}{2}-2\sin
^{2}\theta _{W}\right) \left( N_{p}-N_{e}\right) -\frac{1}{2}N_{n}\right] +%
\tilde{b}_{\nu _{_{\mu ,\tau }}}, \\
c_{\nu _{e}}& =2\sqrt{2}G_{F}\left[ g_{A}\left( 1+2m_{p}\frac{\kappa _{p}}{e}%
\right) C_{p}-g_{A}2m_{n}\frac{\kappa _{n}}{e}C_{n}-C_{e}\right] , \\
c_{\nu _{\mu ,\tau }}& =2\sqrt{2}G_{F}\left[ g_{A}\left( 1+2m_{p}\frac{%
\kappa _{p}}{e}\right) C_{p}-g_{A}2m_{n}\frac{\kappa _{n}}{e}C_{n}+C_{e}%
\right] ,
\end{align}%
where $e>0$ is the proton charge, $\kappa _{n,p}$ are the anomalous part of
the nucleon magnetic moments%
\begin{equation}
\kappa _{n}=-1.91\frac{e}{2m_{n}},\ \ \ \ \ \kappa _{p}=1.79\frac{e}{2m_{p}},
\end{equation}%
and
\begin{eqnarray}
\tilde{b}_{\nu _{l}} &=&\sqrt{2}G_{F}\left[ N_{\nu _{l}}-N_{\bar{\nu}%
_{l}}+\sum_{_{l^{\prime }}}\left( N_{\nu _{l^{\prime }}}-N_{\bar{\nu}%
_{l^{\prime }}}\right) \right] , \\
C_{e,n,p} &=&\frac{1}{2}\int \frac{d^{3}p}{(2\pi )^{3}}\frac{1}{2E}\frac{d}{%
dE}f_{e,n,p}\ .
\end{eqnarray}%
In these equations $N_{e}$ and $N_{\nu _{l}}$($N_{\bar{\nu}_{l}}$) are the
number densities of electrons and neutrinos (antineutrinos) of flavor $l$,
respectively and $f_{e,n,p}$ are the distribution functions of electrons,
protons, and neutrons. Since sterile neutrinos do not interact with the
medium%
\begin{equation*}
V_{\nu _{s}}=V_{\bar{\nu}_{s}}=0,
\end{equation*}%
which means that, unlike the oscillations among active neutrinos, the
neutral current contributions to $V_{\nu _{l}}$ ($V_{\bar{\nu}_{l}})$ play a
relevant role in the MSW equation governing the active-sterile oscillations
in matter.

For antineutrinos the potentials change in sign, i.e. $V_{\bar{\nu}%
_{l}}=-V_{\nu _{l}}$, and in general the transition probability for $\nu
_{l}\longleftrightarrow \nu _{s}$ will be different for neutrinos and
antineutrinos. As a consequence, their relative amounts will change, driving
the potential locally to zero and the mixing angle in matter close to the
one in vacuum throughout the protoneutron star\cite{fuller}. In this work we
will assume that this effect is not important because the resonance
conversion takes place within a thin layer. Under such conditions the kick
mechanism happens as described in Ref. \cite{KS1}.

For relativistic degenerate electrons $C_{e}=-\mu _{e}/8\pi ^{2}$ and for
classical nonrelativistic nucleons $C_{n,p}=-N_{n,p}/8Tm_{n,p}$\cite{nu}\cite%
{DN}, which gives%
\begin{eqnarray}
c_{\nu _{e}} &\simeq &-4\sqrt{2}G_{F}\ \left( \frac{N_{p}}{5Tm_{n}}+\frac{%
3N_{n}}{20Tm_{n}}-\frac{\mu _{e}}{16\pi ^{2}}\right) ,  \label{ce} \\
c_{\nu _{\mu ,\tau }} &\simeq &-4\sqrt{2}G_{F}\ \left( \frac{N_{p}}{5Tm_{n}}+%
\frac{3N_{n}}{20Tm_{n}}+\frac{\mu _{e}}{16\pi ^{2}}\right) ,  \label{cmt}
\end{eqnarray}%
where we have taken $m_{p}\simeq m_{n}$. Here $T$ is the background
temperature and $\mu _{e}=\left( 3\pi ^{2}N_{e}\right) ^{1/3}$ is the
electron chemical potential. It should be noticed that in Ref. \cite{KSS} $%
c_{\nu _{e}}$vanishes because all the components of the stellar plasma are
assumed to be degenerated, and the contributions due to the anomalous
magnetic moment of the nucleons are not included. These expressions have
been derive in the weak-field limit , i.e. $B\ll \mu _{e}^{2}/2e$\cite{nu}.
More general features of the neutrino propagation in magnetized media,
incorporating the effect of strong magnetic fields, have been considered by
several authors\cite{sem}.

The neutrino fractions are much smaller than the electron fraction $%
Y_{e}=N_{e}/(N_{n}+N_{p})$ $\sim 0.1$ and, in the context of our analysis,
the contributions to $V_{\nu _{l}}$ coming from the neutrino-neutrino
interactions (denoted by $\tilde{b}_{\nu _{l}}$) can be neglected. With this
approximation in mind, and taking into account that by electrical neutrality
$N_{e}=$ $N_{p}$, the coefficients in the effective potentials become:
\begin{eqnarray}
b_{\nu _{l}} &=&-b_{\bar{\nu}_{l}}\simeq -\frac{G_{F}}{\sqrt{2}}\left(
1-\eta _{l}Y_{e}\right) \frac{\rho }{m_{n}},  \label{bn} \\
c_{\nu _{l}} &=&-c_{\bar{\nu}_{l}}\simeq -\sqrt{2}G_{F}\left( \frac{3+Y_{e}}{%
5Tm_{n}^{2}}\rho -\xi _{l}\frac{\mu _{e}}{4\pi ^{2}}\right)   \label{cn}
\end{eqnarray}%
where $\eta _{e}=3$, $\xi _{e}=1$ and $\eta _{\mu ,\tau }=1$, $\xi _{\mu
,\tau }=-1$.

As is shown by Eq. (\ref{v2}), the electron fraction plays a very important
role in the case of active-sterile oscillations. This quantity can be
estimated as follows. From the equilibrium condition for the reaction $%
e^{-}+p\leftrightarrows n+\nu _{e}$, taking $\mu _{_{\nu _{e}}}$ $\simeq 0$
we have the relation $\mu _{e}+\mu _{p}\simeq \mu _{n}$ for the chemical
potentials. Hence, $N_{n}=e^{-(Q-\mu _{e})/T}$ $N_{p}$ for a
non-relativistic nucleon gas at a temperature $T$ and
\begin{equation}
Y_{e}\simeq \frac{1}{1+e^{\mu _{e}/T}}\text{ },  \label{ye}
\end{equation}%
where we used the fact that $Q\ll \mu _{e}$ for typical densities in a
protoneutron star. The above is an implicit equation for $Y_{e}$, which has
to be solved numerically once $\rho $ and $T$ are known as a function of $r.$

Now we use the effective potentials with $b_{\bar{\nu}_{l}}$ and $c_{\bar{\nu%
}_{l}}$ given by Eqs. (\ref{bn}) and (\ref{cn}) to discuss the effect of a
magnetic field on the active-sterile matter neutrino oscillations. To make
our analysis simpler, let us consider the case of two states ($\nu _{l}$, $%
\nu _{s}$) that are a mixture of the mass eigenstates $\nu _{1}$ and $\nu
_{2}$. For a momentum $\mathbf{k}$, a $\nu _{l}$ will undergo a resonant
conversion into in a $\nu _{s}$ if the following condition is satisfied:
\begin{equation}
\frac{\Delta m^{2}}{2k}\cos 2\theta =\left. V_{\nu _{l}}\right\vert _{r=R(%
\mathbf{k})},  \label{deltam}
\end{equation}%
where $\theta $ is the vacuum mixing angle and $\Delta m^{2}$ $=$ $m_{2}^{2}$
$-$ $m_{1}^{2}$ is the difference between the squares of the neutrino
masses. Assuming $\Delta m^{2}\cos 2\theta >0$, if $B=0$ the resonant
condition for the electron neutrino (antineutrino) is satisfied when $%
Y_{e}>1/3$ ($Y_{e}<1/3$). In addition, since $Y_{e}<$ $1$ only the $\bar{\nu}%
_{\mu ,\tau }$ (but not the $\nu _{\mu ,\tau }$) can go through the
resonance region. In what follows we assume that within the protoneutron
star $Y_{e}<1/3$, so that the resonant condition is verified only for
antineutrinos.\

In a linear approximation, we can write
\begin{equation}
R(\mathbf{k})=R_{r}(k)+\delta (k)\ \mathbf{\hat{k}}\cdot \mathbf{\hat{B}},
\label{rk}
\end{equation}%
where $R_{r}(k)$ is the radius of the resonance sphere in the absence of the
magnetic field. For a certain value of the magnitude of the momentum $k$,
Eq. (\ref{rk}) determines a spherical shell limited by the spheres of radii $%
R_{r}(k)\pm \delta (k)B,\ $where the resonance condition is verified for
antineutrinos moving in different directions with respect to $\mathbf{B.}$
Using the expression (\ref{v2}) for $V_{\bar{\nu}_{l}}$ in Eq. (\ref{deltam}%
), we get
\begin{equation}
\delta _{\bar{\nu}_{l}}(k)=\mathcal{D}_{\bar{\nu}_{l}}eB,  \label{delta}
\end{equation}%
where%
\begin{equation}
\mathcal{D}_{\bar{\nu}_{l}}=\left. \frac{1}{h_{b_{\bar{\nu}_{l}}}^{-1}}\frac{%
c_{\bar{\nu}_{l}}}{b_{\bar{\nu}_{l}}}\right\vert _{r=R_{r}(k)},  \label{d}
\end{equation}%
and we have defined $h_{g}^{-1}\equiv \frac{d}{dr}\ln g$ for any function $%
g(r)$. From Eqs. (\ref{bn}) and (\ref{cn}) we find explicitly
\begin{equation}
\mathcal{D}_{\bar{\nu}_{l}}=\left. \frac{2}{\left( 1-\eta _{l}Y_{e}\right)
h_{\rho }^{-1}-\eta _{l}Y_{e}h_{Y_{e}}^{-1}}\left( \frac{3+Y_{e}}{5Tm_{n}}%
-\xi _{l}\frac{m_{n}\mu _{e}}{4\pi ^{2}\rho }\right) \right\vert
_{r=R_{r}(k)}.  \label{d2}
\end{equation}%
By means of Eq. (\ref{ye}), $h_{Y_{e}}^{-1}$ can be expressed in terms of $%
h_{\rho }^{-1}$ and $h_{T}^{-1}$ as follows%
\begin{equation}
h_{Y_{e}}^{-1}=-\left( h_{\rho }^{-1}-3h_{T}^{-1}\right) \frac{1-Y_{e}}{%
1-Y_{e}+\frac{3T}{\mu _{e}}}.  \label{hy}
\end{equation}

The efficiency of the flavor transformation depends on the adiabaticity of
the process. The conversion will be adiabatic when the oscillation length at
the resonance is smaller than the width of the enhancement region. This
imposes the condition%
\begin{equation}
\left\vert \frac{dV_{\bar{\nu}_{l}}}{dr}\right\vert _{r=R_{r}(k)}<\left.
\left( \frac{\Delta m^{2}}{2k}\sin 2\theta \right) ^{2}\right\vert
_{r=R_{r}(k)}.  \label{hh}
\end{equation}%
From Eq. (\ref{v2}) and neglecting the contribution of the magnetic field we
get%
\begin{equation}
\tan ^{2}2\theta >\left\vert \frac{1}{b_{\bar{\nu}_{l}}}\left( h_{\rho
}^{-1}-\frac{\eta _{l}Y_{e}}{1-\eta _{l}Y_{e}}h_{Y_{e}}^{-1}\right)
\right\vert _{r=R_{r}(k)},  \label{ac}
\end{equation}%
with $b_{\bar{\nu}_{l}}$ given by Eq. (\ref{bn}).

\section{Neutrino momentum asymmetry}

The energy flux in the atmosphere is radial and determined by the diffusive
part of the neutrino distribution function. Consequently, in previous works
on the subject the resonance condition has been evaluated in terms of an
average neutrino momentum $\left\langle k\right\rangle =\frac{7\pi ^{4}}{%
180\zeta (3)}T\simeq 3.15T$ pointing in the radial direction. This procedure
leads to the concept of a deformed average surface of resonance, which acts
as the boundary of an effective neutrinosphere. In a more detailed
description, at each point of the neutrinosphere we have neutrinos with
momentum $\mathbf{k}$ pointing in all directions. Hence, the resonance
condition as given by Eq. (\ref{deltam}) actually defines a spherical
surface for each value of $\mathbf{\hat{k}\cdot \hat{B}}$, which acts as a
source of sterile neutrinos moving in the $\mathbf{\hat{k}}$ direction.
Accordingly, as we mentioned above, for a given $k$ the resonant transition
takes place within a spherical shell instead of a deformed spherical
surface. We will discuss the situation where the resonance shell lies
totally within the $\bar{\nu}_{l}$-neutrinosphere. If the mixing angle is
small and the adiabatic condition of Eq. (\ref{ac}) is satisfied, then the $%
\bar{\nu}_{l}$ are almost completely converted into $\bar{\nu}_{s}$. The
outgoing sterile antineutrinos freely escape from the protostar, while the
ingoing ones propagate through the medium until they cross again the
resonance surface and are reconverted into $\bar{\nu}_{l}$, which are
thermalized. In consequence, only those $\bar{\nu}_{s}$ going outward leave
the star and the resonance surfaces for each momentum direction act as
effective emission semispheres. For $\mathbf{k}$ pointing in opposite
directions the respective semispheres are at $R_{r}(k)\pm \delta (k)kB$ and
therefore have different areas, which generates a difference in the momentum
carried away by the sterile neutrinos leaving the star in opposite
directions. This difference adds up to building a nonvanishing asymmetry in
the total momentum $\mathcal{K}$ emitted by the cooling neutron star, which
we now compute explicitly in terms of the whole neutrino distribution
function.

Within the neutrinosphere the $\bar{\nu}_{l}$ satisfy the diffusion regime%
\cite{BDMZ}\cite{ss} with a distribution function%
\begin{equation}
f_{_{\bar{\nu}_{l}}}=f_{_{\bar{\nu}_{l}}}^{eq}-\frac{1}{\kappa _{\bar{\nu}%
_{l}}\rho k^{3}}\mathbf{k}\cdot \mathbf{\nabla }f_{_{\bar{\nu}_{l}}}^{eq},
\label{distri}
\end{equation}%
where $f_{_{_{\bar{\nu}_{l}}}}^{eq}=\left( 1+e^{k/T}\right) ^{-1}$ is the
distribution function at equilibrium. We have again assumed that the
neutrino chemical potential is negligible. The momentum flux emitted in the
radial direction is%
\begin{equation}
dF_{r}=\frac{\mathbf{\hat{k}\cdot \hat{r}}}{(2\pi )^{3}}f_{_{_{\bar{\nu}%
_{l}}}}kd^{3}k.  \label{flux}
\end{equation}%
Let us now adopt a reference frame ($x,y,z$) fixed to the star, where the
magnetic field coincides with the $z$-axis, $\mathbf{B}=B\mathbf{\hat{z}}$.
In addition, at each point within the spherical shell determined by Eq. (\ref%
{rk}) we use a local frame ($x^{\prime },y^{\prime },z^{\prime }$) with $%
\mathbf{\hat{r}}=\mathbf{\hat{z}}^{\prime }$, such that $\mathbf{\hat{k}}$ $%
=\cos \theta ^{\prime }\mathbf{\hat{z}}^{\prime }+\sin \theta ^{\prime }\cos
\phi ^{\prime }\mathbf{\hat{x}}^{\prime }$ $+\sin \theta ^{\prime }\sin \phi
^{\prime }\mathbf{\hat{y}}^{\prime }$. In this way the total momentum
carried away by the $\bar{\nu}_{l}$ can be expressed as%
\begin{eqnarray}
k_{r} &=&\frac{1}{(2\pi )^{2}}\int_{0}^{\infty }k^{3}dk\int_{0}^{\pi }\sin
\theta d\theta \int_{0}^{\frac{\pi }{2}}\sin \theta ^{\prime }d\theta
^{\prime }\int_{0}^{2\pi }d\phi ^{\prime }R^{2}(\mathbf{k})f_{_{_{\bar{\nu}%
_{l}}}}\mathbf{\hat{k}\cdot \hat{r}} \\
&\mathbf{=}&\frac{1}{2\pi }\int_{0}^{\infty }dk\ k^{3}\ R_{r}^{2}(k)\left(
f_{_{_{\bar{\nu}_{l}}}}^{eq}-\frac{2}{3\kappa _{_{\bar{\nu}_{l}}}\rho k^{2}}%
\frac{df_{_{_{\bar{\nu}_{l}}}}^{eq}}{dr}\right) .  \label{er}
\end{eqnarray}%
The momentum emitted in direction of the magnetic field $k_{B}$ is
calculated in a similar manner, with $\mathbf{\hat{k}\cdot \hat{r}}$
replaced by $\mathbf{\hat{k}\cdot \hat{B}}=\cos \theta \cos \theta ^{\prime
}-\sin \theta \sin \theta ^{\prime }\sin \phi ^{\prime }$:%
\begin{equation}
k_{B}=\frac{2}{3\pi }\int_{0}^{\infty }dk\ k^{3}R_{r}(k)\delta (k)\left(
f_{_{_{\bar{\nu}_{l}}}}^{eq}-\frac{1}{2\kappa _{_{\bar{\nu}_{l}}}\rho k^{2}}%
\frac{df_{_{_{\bar{\nu}_{l}}}}^{eq}}{dr}\right) ,  \label{eb}
\end{equation}%
where we have used Eq. (\ref{rk}) keeping at most terms that are linear in $%
\delta (k)$.

In terms of the above quantities the $\bar{\nu}_{l}$ contribution to the
fractional asymmetry in the total momentum becomes%
\begin{equation}
\frac{\Delta \mathcal{K}}{\mathcal{K}}=\frac{1}{6}\frac{k_{B}}{k_{r}},
\label{momas}
\end{equation}%
where the factor $1/6$ comes from the fact that we are assuming that $%
\mathcal{K}$ is equipartitioned among all the neutrino and antineutrino
flavors. In order to evaluate the remaining integrals in Eqs. (\ref{er}) and
(\ref{eb}) we need to know the explicit dependence on $k$ of the functions $%
R_{r}$ and $\delta $. Simple analytical results can be derived by replacing
these functions by their values at the average momentum $\left\langle
k\right\rangle .$ Proceeding in this manner we get
\begin{eqnarray}
k_{r} &=&\frac{\pi R_{r}^{2}}{48}\left( \frac{7\pi ^{2}T^{4}}{5}-\frac{4}{%
3\kappa _{_{\bar{\nu}_{l}}}\rho }\frac{dT^{2}}{dr}\right) ,  \label{erad} \\
k_{B} &=&\frac{\pi R_{r}\delta }{36}\left( \frac{7\pi ^{2}T^{4}}{5}-\frac{1}{%
\kappa _{_{\bar{\nu}_{l}}}\rho }\frac{dT^{2}}{dr}\right) .  \label{brad}
\end{eqnarray}%
In these expressions the terms that depend on $dT^{2}/dr$ come from the
diffusive part of the distribution function. In the regime considered here
they happen to be smaller than the first terms and we will discard them.
This can be verified in the context of specific models like the one
considered in the next section. Using this approximation, we finally obtain
\begin{equation}
\frac{\Delta \mathcal{K}}{\mathcal{K}}\simeq \frac{2}{9}\frac{\delta }{R_{r}}%
,  \label{deltk}
\end{equation}%
which is four times bigger than the asymmetry we derived by assuming a
single (deformed) resonance surface\cite{BDMZ}. This means that if Eq. (\ref%
{deltk}) were applied to the case of active-active oscillations, when the
resonance shell is at the border of the electron neutrinosphere, we would
obtain $B\simeq 7.5\times 10^{15}\ G$ for the intensity of the magnetic
field required to produce the observed pulsar velocities.

Up to this point we have considered the resonant oscillation between a
sterile antineutrino and an active antineutrino of a given flavor. The
observational evidence indicates that active neutrinos mix. Furthermore, it
is reasonable to assume that the three active antineutrinos could be mixed
with the sterile ones. Consequently, the situation can be more complicated
than the simple pairwise neutrino oscillation already discussed. In general
different mixing schemes will render gives different results for the
magnetic field, but these results will differ in factors of the order of the
unity. Therefore, to make an estimation of the magnetic field we will
consider the simplest possible scheme, in which the mixing between active
neutrinos is neglected and pairwise oscillation with the sterile ones are
taken. In this context the total momentum asymmetry can be written%
\begin{equation}
\frac{\Delta \mathcal{K}}{\mathcal{K}}\simeq \frac{2}{9}\left( \frac{\delta
_{\bar{\nu}_{e}}}{R_{r_{\bar{\nu}_{e}}}}+2\frac{\delta _{\bar{\nu}_{\mu
,\tau }}}{R_{r_{\bar{\nu}_{\mu ,\tau }}}}\right)  \label{deltkk}
\end{equation}%
In what follows we will make the calculation for the active-sterile
conversion in the framework of this approximation by assuming that all the
resonance layers are entirely within the $\nu _{\mu ,\tau }$-neutrinosphere.

\section{Numerical results}

Taking into account the total momentum emitted by the protostar, to have the
required kick $\Delta \mathcal{K}$/$\mathcal{K}$ must be of the order of $%
0.01$. Inserting this value for the momentum asymmetry on the left hand side
of Eq. (\ref{deltkk}) we see that $\delta _{\bar{\nu}_{l}}\sim 0.015R_{r_{%
\bar{\nu}_{l}}}$, which means that the thickness of the resonance shell is
of the order of one hundred of meters. From Eqs. (\ref{delta}) and (\ref%
{deltkk}), using $R_{r_{\bar{\nu}_{e}}}\simeq R_{r_{\bar{\nu}_{\mu ,\tau
}}}\simeq R_{r}$, we find%
\begin{equation}
eB=0.045\frac{R_{r}}{\mathcal{D}_{\bar{\nu}_{e}}+2\mathcal{D}_{\bar{\nu}%
_{\mu ,\tau }}}  \label{bedd}
\end{equation}%
The magnitude of the magnetic field needed to produce the kick can be
evaluated from this result and Eq. (\ref{d2}), once the temperature and
density profiles are known. In this section we will perform this calculation
for two simple models of a protoneutron star atmosphere. One of them is the
spherical Eddington model\cite{BDMZ} and the other is the one introduced in
Ref. \cite{BM}, on the basis of a fitting of numerical results.

The Eddington model\cite{ss} gives a satisfactory description of a
neutrinosphere locally homogeneous and isotropic, with a conserved energy
flux. In the case of a protoneutron star the results are in reasonable
agreement with those obtained by numerical analysis. For a spherical
geometry the model was developed in Ref. \cite{BDMZ}, where the reader is
referred for details. Here we summarize those features that are relevant for
our analysis. The atmosphere of the protostar is described as an ideal
nucleon gas in hydrostatic equilibrium%
\begin{eqnarray}
P &=&\frac{\rho T}{m_{n}},  \label{e1} \\
\frac{dP}{dr} &=&-\rho \frac{GM(r)}{r^{2}},  \label{e2}
\end{eqnarray}%
with $M(r)=4\pi \int_{0}^{r}dr^{\prime }\;r^{\prime 2}\rho (r^{\prime })$.
The total neutrino momentum flux satisfies the diffusion equation
\begin{equation}
F(r)=-\frac{1}{36}\frac{1}{\bar{\kappa}\rho (r)}\frac{d}{dr}T^{2}(r),
\label{f}
\end{equation}%
where $\bar{\kappa}=\left( \kappa _{_{\bar{\nu}_{e}}}^{-1}+2\kappa _{_{\bar{%
\nu}_{\mu ,\tau }}}^{-1}\right) ^{-1}=3\times 10^{-26}\ MeV^{-5}$, and the
flux conservation
\begin{equation}
F(r)=\frac{L_{c}}{4\pi r^{2}},  \label{e5}
\end{equation}%
where $L_{c}$ is the luminosity of the protostar. Eq. (\ref{f}) is obtained
from Eq. (\ref{flux}) by integrating over the momentum.

The above system of equations has no exact analytical solution. However, a
good approximate solution has been found in Ref. \cite{BDMZ}. It allows us a
consistent treatment, from which we obtain the following expressions:%
\begin{equation}
h_{T}^{-1}=-\frac{\lambda _{c}}{1-a}\frac{R_{c}}{r^{2}}\frac{T_{c}}{T}\left(
1-\frac{T_{c}^{2}}{T^{2}}a\right) ,  \label{dif}
\end{equation}%
\begin{equation}
h_{\rho }^{-1}+h_{T}^{-1}=-\frac{Gm_{n}M(r)}{r^{2}T},  \label{ht}
\end{equation}%
with $\lambda _{c}=(9\kappa L_{c}\rho _{c})/(2\pi T_{c}^{2}R_{c})$. Here $%
\rho _{c}$ and $T_{c}$ are the density and the temperature at the radius of
the core $R_{c}$, and $a$ is a function of the temperature
\begin{equation}
a=1-\frac{1}{\alpha _{c}}-\frac{T_{c}-T}{T_{c}-T_{s}}\left( 1-\frac{1}{%
\alpha _{c}}-\frac{T_{s}^{2}}{T_{c}^{2}}\right) \,,  \label{at}
\end{equation}%
where $\alpha _{c}=(GM_{c}m_{n})/(2\lambda _{c}T_{c}R_{c})$ and $T_{s}$ is
the asymptotic value of the temperature. In Fig. 1 we show the resulting
profiles for $\rho (r),$ $T(r)$, and $M(r)$, corresponding to a
configuration like the one considered in Ref. \cite{BDMZ}: $M_{c}=M_{\odot }$%
, $R_{c}=10\ km$, $L_{c}=2\times 10^{52}\;erg\;s^{-1}$, $\rho
_{c}=10^{14}\;g\ cm^{-3}$, $T_{c}=40\ MeV$, and $T_{s}=$ $4.8$ \ $MeV$. In
the same figure, we plot $Y_{e}(r)$ as obtained by solving Eq. (\ref{ye}) at
every point in the atmosphere.

According to Eq. (\ref{ln}) the electron neutrino mean free path is
\begin{equation}
\lambda _{\bar{\nu}_{e}}\simeq 8\left( \frac{\rho _{c}}{\rho }\right) \left(
\frac{T_{c}}{T}\right) ^{2}\ cm.  \label{fp1}
\end{equation}%
The radius of the neutrinosphere is estimated by inserting the above
expression into Eq. (\ref{nc}) and evaluating the resulting integral
numerically, the result is $R_{\bar{\nu}_{e}}\simeq 2.6\ R_{c}$. This yields
$\rho (R_{\bar{\nu}_{e}})\simeq 1.3\times 10^{11}\ g\ cm^{-3}$, $T(R_{\bar{%
\nu}_{e}})\simeq T_{s}$ , and $Y_{e}(R_{\bar{\nu}_{e}})\simeq 0.1$ which are
in good agreement with typical values for the electron neutrinosphere in a
protoneutron star. In the same way, for the muon (tau) neutrino we get $R_{%
\bar{\nu}_{\mu ,\tau }}\simeq 2.2\ R_{c}$.

Using Eqs. (\ref{e5}) and (\ref{f}), the magnitude of the ratio of the
second and the first terms in Eq. (\ref{erad}) is $(L_{c}\kappa _{\nu
_{l}})/(3\bar{\kappa}T^{4}r^{2})$, which varies between $10^{-4}$ and $%
10^{-1}$ for $R_{c}\leqslant r\leqslant R_{\bar{\nu}_{\mu ,\tau }}$.
Therefore, the approximation done in writing Eq. (\ref{deltk}) is justified
and the intensity of the magnetic field required to produce the pulsar kick
can be evaluated from Eq. (\ref{bedd}). As shown in Fig. 2, at the surface
of the core $B_{c}\simeq 8\times 10^{16}\ G$ and $B$ remains approximately
constant when the position of the resonance moves through the protostar
atmosphere. According to Eq. (\ref{deltam}), in the interval $R_{c}\
\leqslant R_{r}\leqslant R_{_{\bar{\nu}_{\mu ,\tau }}}$ the allowed values
of the oscillations parameters are $10^{9}\ eV^{2}\gtrsim \Delta m^{2}\cos
2\theta \gtrsim 10^{6}\ eV^{2}$, that for small mixing leads to $30\
KeV\gtrsim $ $m_{s}\gtrsim 1\ KeV$. From Eq. (\ref{ac}) it can be seen that
transitions will be adiabatic for $\tan ^{2}2\theta >7\times 10^{-12}$ at $%
R_{r}=R_{c}$ and $\tan ^{2}2\theta >3\times 10^{-9}$ at $R_{r}=R_{_{\bar{\nu}%
_{\mu ,\tau }}}$.

In the second model, both the baryon density and the temperature profiles
follows simple potential laws and are related according to \cite{BM}%
\begin{equation}
\rho =\rho _{c}\left( \frac{T}{T_{c}}\right) ^{3},
\end{equation}%
From this expression we see that $h_{\rho }^{-1}=3h_{T}^{-1}$, which in turn
implies that $h_{Y_{e}}^{-1}=$ $0$ (see Eq. (\ref{hy})). As a consequence,
the electron fraction $Y_{e}$ and also the ratio $\mu _{e}/T$ remain
constants in the protostar atmosphere. Employing the same set of core
parameters than in the preceding model we have $Y_{e}=0.08$ and $\mu
_{e}/T=2.5$. In what follows, for the density profile we use $\rho =\rho
_{c}\left( \frac{R_{c}}{r}\right) ^{4}$, which gives $h_{\rho }^{-1}=$ $-4/r$
, and the border of the $\bar{\nu}_{\mu ,\tau }$-neutrinoshere is located at
$\rho (R_{\bar{\nu}_{\mu ,\tau }})\simeq 10^{12}\ g\ cm^{-3}$, that
corresponds to $R_{\bar{\nu}_{\mu ,\tau }}\simeq 3\ R_{c}$ and $T(R_{\bar{\nu%
}_{\mu ,\tau }})\simeq 9\ MeV$. The ratio of the diffusive term to the
isotropic term in Eq. (\ref{erad}) can be expressed as $10^{-4}\left(
r/R_{c}\right) ^{17/3}$ and the approximation (\ref{deltk}) for $\Delta
\mathcal{K}/\mathcal{K}$ is also valid in this model. Taking these results
into account and, from Eqs. (\ref{d}) and (\ref{bedd}) the following simple
expression can be derived for the required magnetic field:%
\begin{equation}
B=3\times 10^{17}\left( 1+0.09\frac{T_{c}}{T}\right) ^{-1}\frac{T}{T_{c}}\ G,
\end{equation}
with $T=$ $T_{c}(R_{c}/r)^{4/3}$. Now the intensity of the magnetic field
decreases monotonically from $3\times 10^{17}\ G$ at $R_{r}=R_{c}$ to $%
5\times 10^{16}\ G$ at $R_{r}=R_{\bar{\nu}_{\mu ,\tau }}$. The allowed range
for the sterile neutrino mass is similar to the one derived in the case of
the Eddington model, while the requirement that the resonant conversion be
adiabatic imposes the condition $\tan ^{2}2\theta \gtrsim 3\times
10^{-11}\left( R_{r}/R_{c}\right) ^{3}$ on the mixing angle.

A comment is in order. Our results have been derived on the basis of the
weak field limit for the effective neutrino potentials in a magnetized
medium. As it has been already mentioned in Section 2, this approximation is
valid whenever $B\ll \mu _{e}^{2}/2e$. In one of the examples (the Eddington
model) this constraint is verified in the interior but not in the external
region of the protostar. A more careful treatment would require to
incorporate the effect of strong magnetic fields along the lines examined in
Ref. \cite{nu}.

\section{Final remarks}

We have reexamined the anisotropic momentum emission by a nascent neutron
star driven by the resonant neutrino conversion in presence of strong
magnetic fields. The protostar atmosphere was modeled by a nonrelativistic
nucleon gas and a degenerated electron gas. We elaborate the concept of a
resonance spherical shell for neutrinos with a given momentum $k$ moving in
different directions relative to $\mathbf{B}$. This concept provides a more
precise description of the problem than a deformed spherical surface for
neutrinos with an average radial momentum. An important ingredient for the
existence of the effect is that the ingoing neutrinos become thermalized.
For active-active neutrino oscillations ($\nu _{e}\longleftrightarrow $ $\nu
_{\mu ,\tau }$) the thermalization proceeds by any of the following reasons:
either because the ingoing $\nu _{\mu ,\tau }$ moving toward the interior of
the protoneutron star enter their own neutrinosphere, or because they cross
again the resonance region and are reconverted into $\nu _{e}$, which are
now within their neutrinosphere. For sterile neutrinos solely the second
process is effective. In any case only the outgoing neutrinos contribute to
the total momentum flux . The difference in the areas of the semispherical
emission surface for neutrinos moving in opposite directions with the same
momentum magnitude originates a non null fractional asymmetry in the total
momentum. We apply our approach to the study of the generation of a natal
pulsar kick through the $\bar{\nu}_{l}\longleftrightarrow \bar{\nu}_{s}$
oscillations. The kick was calculated in a two simple model for the stellar
atmosphere and in both cases a magnetic field of the order of $10^{17}\ G$
is required to produce the observed velocities. The ranges of the mass and
the mixing angle are compatible with a massive sterile neutrino as a warm
dark matter candidate\cite{dh}. Fields of this magnitude are possible in
protoneutron stars. For example, magnetic fields of the order of $10^{18}\ G$
have been considered in the literature for the central regions\cite{fermi},
with a dependence in the baryon density parametrized as\cite{manka}
\begin{equation}
B_{PNS}=B_{s}+B_{c}\left[ 1-e^{-\beta \left( \rho /\rho _{s}\right) ^{\gamma
}}\right] ,  \label{intb}
\end{equation}%
with $\beta =10^{-5}$ and $\gamma =3.$ In Eq. (\ref{intb}) $\rho
_{s}=10^{11}g\ cm^{-3}$,\ $B_{s}=10^{14}\ G$, and $B_{c}=10^{18}\ G$ are the
strengths of the magnetic field at the surface and the core, respectively.
For comparison, in Fig. 2 and 3 we have also shown the magnetic field as
given by Eq. (\ref{intb}). Taking this curve as an upper value of the
magnetic field in the protostar, we see that the resonant neutrino
transformation could be an effective mechanism for the pulsar kick in most
of the neutrinosphere.

\section{Acknowledgments}

We thank V. Semikoz and A. Kusenko for useful comments. This work was
partially supported by CONICET-Argentina, CONACYT- M\'{e}xico, and
PAPIIT-UNAM grant IN109001. M. B. also acknowledges support from DGEP-UNAM (M%
\'{e}xico).

\pagebreak

\section*{Figure captions}

FIG. 1: Characteristic profiles for the Eddington model. The different
functions are normalized to the corresponding value at the core: $R_{c}=10\
km$, $T_{c}=40\ MeV$, $\rho _{c}=\ 10^{14}g\ cm^{3}$, $M_{c}=\ 1\ M_{\odot }$%
, and $Y_{ec}=\ 0.08$.

FIG. 2: Magnetic field required to produce the kick, $B$, for the Eddington
model and the expected maximum magnetic field, $B_{PNS}$, as functions of
the resonance radius and normalized to the field required at the core, $%
B_{c}=\ 8\times 10^{16}\ G$.

FIG. 3: Magnetic field required to produce the kick, $B$, for the potential
model and the expected maximum magnetic field, $B_{PNS}$, as functions of
the resonance radius and normalized to the field required at the core, $%
B_{c}=3\times 10^{17}\ G$.

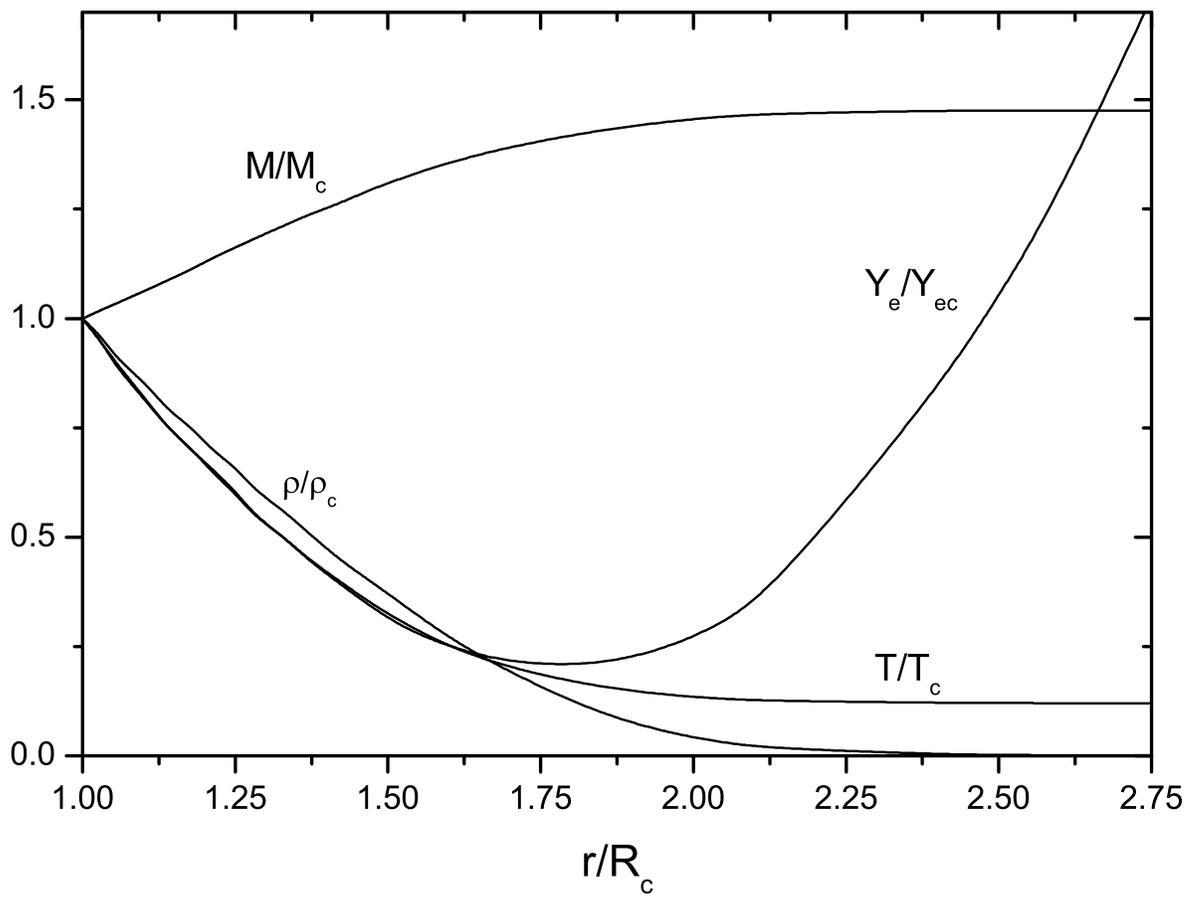
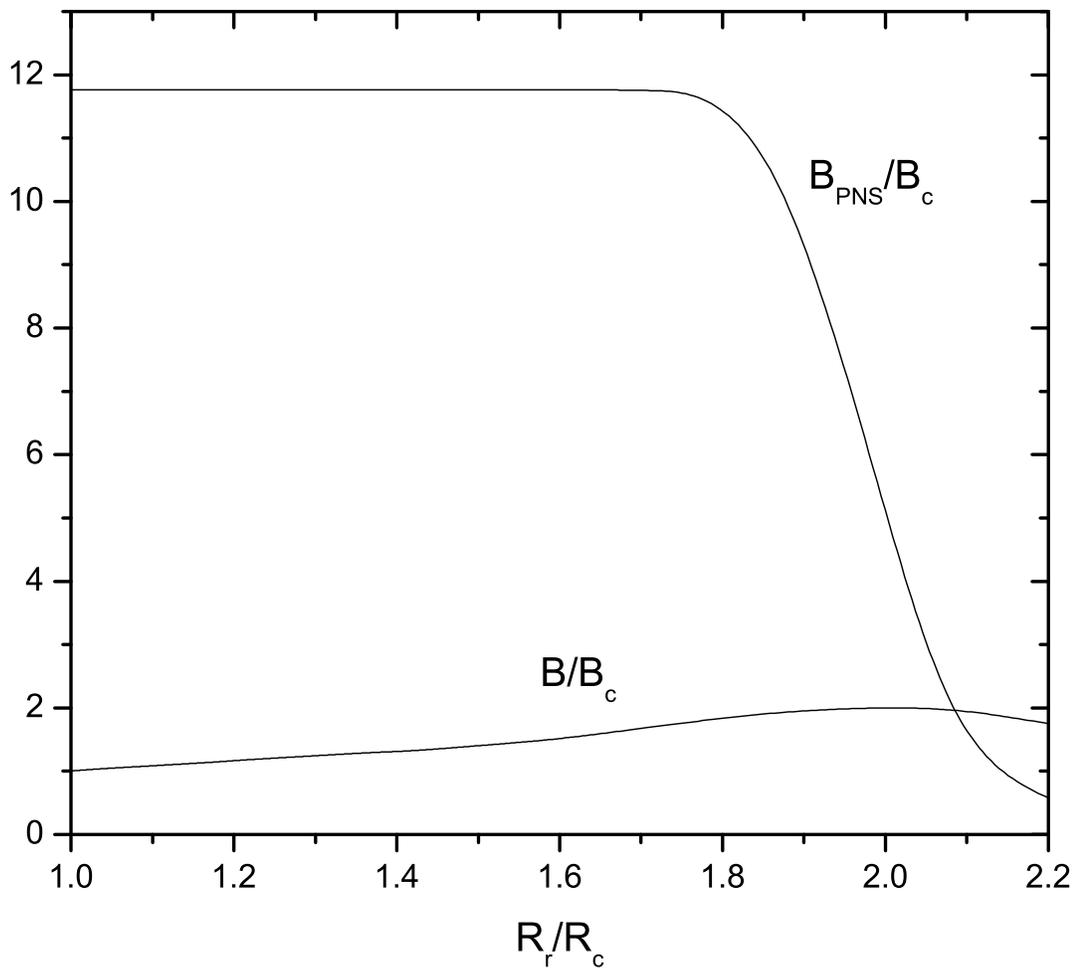
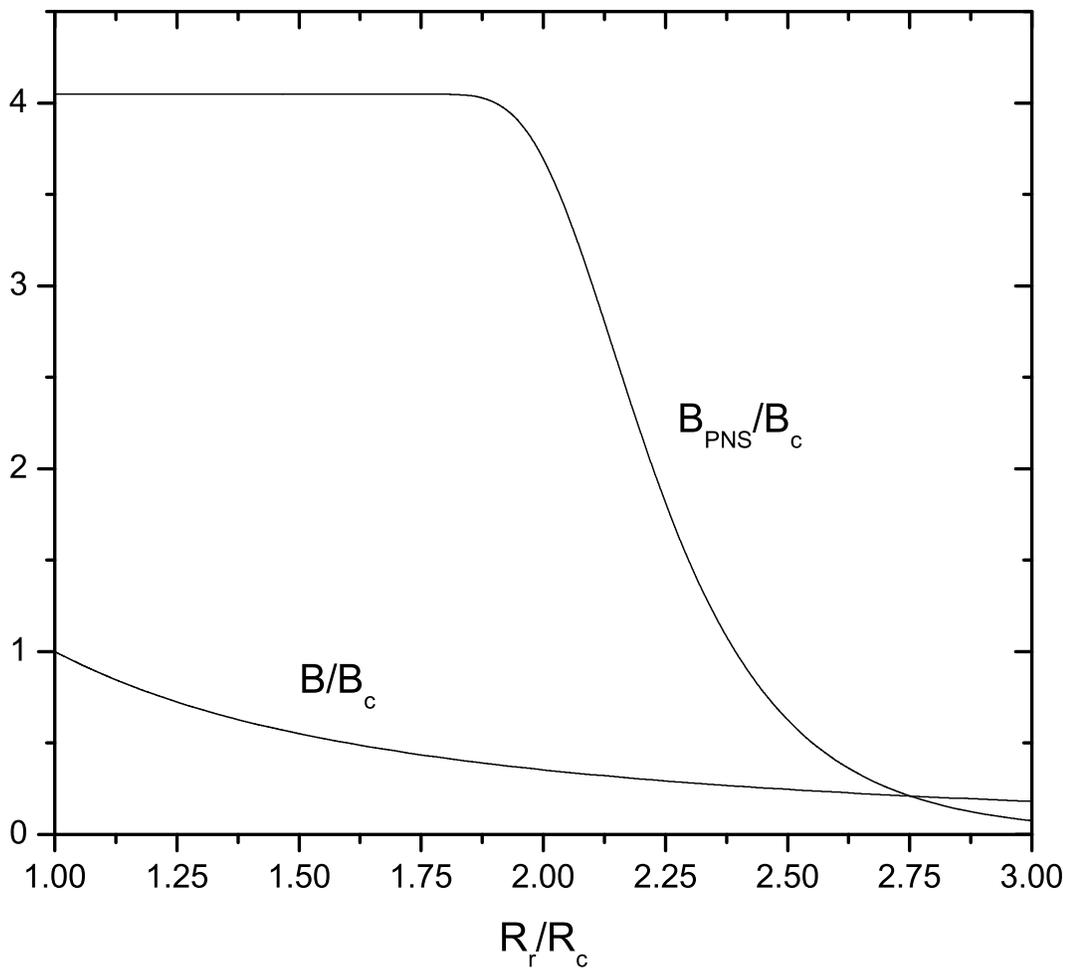
\end{document}